# Force unfolding kinetics of RNA using optical tweezers.
# I. Effects of experimental variables on measured results

Jin-Der Wen,[*] Maria Manosas,[†] Pan T.X. Li,[*] Steven B. Smith,[‡] Carlos Bustamante,[*‡§] Felix Ritort,[†] Ignacio Tinoco, Jr.[*]

[*]Department of Chemistry, [†]Departament de Fisica Fonamental, Universitat de Barcelona, Barcelona, Spain, [‡]Department of Physics, and [§]Department of Molecular and Cell Biology and Howard Hughes Medical Institute, University of California, Berkeley, California 94720

# ABSTRACT

Experimental variables of optical tweezers instrumentation that affect RNA folding/unfolding kinetics were investigated. A model RNA hairpin, P5ab, was attached to two micron-sized beads through hybrid RNA/DNA handles; one bead was trapped by dual-beam lasers and the other was held by a micropipette. Several experimental variables were changed while measuring the unfolding/refolding kinetics, including handle lengths, trap stiffness, and modes of force applied to the molecule. In constant-force mode where the tension applied to the RNA was maintained through feedback control, the measured rate coefficients varied within 40% when the handle lengths were changed by 10 fold (1.1 to 10.2 Kbp); they increased by two- to three-fold when the trap stiffness was lowered to one third (from 0.1 to 0.035 pN/nm). In the passive mode, without feedback control and where the force applied to the RNA varied in response to the end-to-end distance change of the tether, the RNA hopped between a high-force folded-state and a low-force unfolded-state. In this mode, the rates increased up to two-fold with longer handles or softer traps. Overall, the measured rates remained with the same order-of-magnitude over the wide range of conditions studied. In the companion paper (1), we analyze how the measured kinetics parameters differ from the intrinsic molecular rates of the RNA, and thus how to obtain the molecular rates.

**ABBREVIATIONS**

bp, base pair; PCR, polymerase chain reaction; *SNR*, signal-to-noise ratio.

**INTRODUCTION**

Discovery of RNA's increasing roles in many biological processes, such as regulation of gene expression, has stimulated interest in understanding how the RNA folds into native structures to perform its functions. Folding of the RNA is highly hierarchical, i.e., the primary sequence of an RNA molecule forms secondary structural elements through base pairs, which subsequently fold to tertiary domains/structures, usually through long-range interactions (2). Moreover, several domains from a large RNA can fold independently and then assemble into more complex structures (3, 4). RNA folding is strongly affected by environmental factors, including magnesium ions. For example, the *Tetrahymena* ribozyme does not form a stable structure in low $Mg^{2+}$ concentrations, whereas $Mg^{2+}$-stabilized kinetic traps (misfolded species) slow the folding of the RNA in high $Mg^{2+}$ concentrations (5). Kinetically trapped, alternatively folded conformers usually occur *in vitro* during folding of larger RNAs, and they can be thermodynamically stable and never fold into correct structures (6).

RNA folding/unfolding thermodynamics and kinetics are traditionally studied by changing the temperature (7, 8) or denaturant (e.g., urea) concentration (9, 10). These variables can affect the equilibria and rates of the RNA folding reactions. Recently, optical tweezers-based single-molecule techniques (11-13) have introduced another variable—mechanical force—to study RNA folding/unfolding (14, 15). This new approach offers several advantages over the traditional methods. First, mechanical forces are involved in many biological processes, such as opening of RNA hairpins by helicases (16). Second, the progress of the reaction can be followed by a well-defined reaction coordinate (end-to-end distance of the RNA). Finally, an RNA molecule usually traverses intermediate conformations before folding to its native structure, and single-molecule approaches make the detection and characterization of the intermediate states more accessible than bulk methods (17, 18).

To facilitate single-molecule manipulation in a typical optical tweezers unfolding experiment, the RNA of interest is attached to two micron-sized beads through molecular "handles", which are generally double-stranded nucleic acids to physically separate the RNA from the beads and to prevent the interference of the bead surfaces. One bead is held in the optical trap and the other is attached to a micropipette. Kinetics of RNA folding and unfolding is studied by monitoring distance changes between the two beads in response to the applied forces. However, several factors in the experimental setup may affect the measured unfolding/refolding rates of RNAs, as has been shown in a recent report on a 20-bp DNA hairpin whose rates change with the stiffness of the optical trap (19).

Our goals in this study are to systematically investigate the experimental influences on the kinetics of RNA hairpins in a typical optical tweezers experiment, and to analyze the limitation of measurements under such conditions. P5ab, a simple 22-bp RNA hairpin derived from the *Tetrahymena thermophila* ribozyme (20), was used as a model. The folding/unfolding rates of the RNA were measured for different handle lengths (1.1, 3.2, 5.9, and 10.2 Kbp), different stiffness of the optical trap (0.1 and 0.035 pN/nm), and two modes of force control (constant-force and passive modes, see below for details). Signal-to-noise ratios (*SNR*) were calculated as a function of force, extension, and time to validate those measurements. In the companion paper (1), we applied a mesoscopic model to simulate RNA kinetics under comparable conditions. By comparing the results from experiments and theory, we were able to deduce the intrinsic molecular rates, the ideal folding/unfolding rates of the RNA under a fixed force and without flanking handles and beads (1). The current experimental and theoretical data will be helpful for future experimental designs to reduce instrumental influences on measurements of force-unfolding kinetics of RNA or DNA.

**MATERIALS AND METHODS**

**Preparation of RNA and single-molecule constructs**

The DNA sequence corresponding to the P5ab RNA was synthesized (Operon, Huntsville, AL) and cloned into a 10.3 Kbp pREP4 vector (Invitrogen, Carlsbad, CA) between the *Hind* III and *Xho* I sites. Based on the cloned vector, four sets of primers were designed for PCR (polymerase chain reaction) to make different lengths of templates, with a T7 promoter sequence (TAATACGACTCACTATAGGG) (21) at the 5′ end. The lengths of the templates were 1.1, 3.2, 5.9, and 10.2 Kbp, corresponding to positions 33 – 1152, 9356 – 2231, 8019 – 3534, and 5849 – 5754, respectively, of the original pREP4 vector. The inserted P5ab sequence (Figure 1A) located approximately at the center of each template. RNA was produced by in vitro transcription; lengths and integrity of the products were verified by denaturing agarose gel electrophoresis. The RNA was annealed to two corresponding single-stranded DNA, handles A and B, which were respectively complementary to the 5′- and 3′-end halves of the RNA transcripts, leaving only the middle P5ab sequence unhybridized (see Figure 1A). The annealing reaction was carried out with approximately equal molar concentrations of RNA and each of the DNA handles in the annealing buffer (64% formamide, 32 mM PIPES, pH 6.3, 320 mM NaCl, and 1 mM EDTA). The mixture was incubated at 85°C for 10 min, 62°C for 90 min, 52°C for 90 min, and ramped to 10°C over 10 min. The hybrid constructs were recovered by ethanol precipitation. The annealing efficiency of RNA to DNA handles is usually difficult to estimate from the gel. We empirically assessed the efficiency of annealing by determining what dilutions of the constructs gave sufficient tethers to beads during tweezers experiments. In this respect, the annealing efficiency for each construct (from 1.1 to 10.2 KB) was similar. The DNA strands of the handles were generated by PCR; handle A was subsequently biotinylated at the 3′ end by an exchange reaction using T4 DNA polymerase (22), whereas a digoxigenin group was introduced at the 5′ end of handle B via the primer during PCR. The biotin and digoxigenin tags on opposite ends of the RNA hybrids provide affinity binding of the constructs to surface-modified polystyrene beads to allow single-molecule manipulation (see below).

**Optical tweezers setup**

The single-molecule manipulation of RNA was done on a force-measuring dual-beam optical tweezers (12, 23). The P5ab RNA was held between two polystyrene beads (Spherotech, Libertyville, IL) by immobilizing the free ends of hybrid RNA/DNA handles onto the surface of the beads. One bead (~ 3 µm in diameter) was cross-linked with anti-digoxigenin antibody and trapped by the lasers; the other (~ 2 µm in diameter) was coated with streptavidin and positioned by suction on the tip of a micropipette, which was fixed in the reaction chamber and coupled to a piezoelectric flexure stage for small displacements. The bead in the laser trap followed Hooke's law, such that the exerted force (measured by changes in light momentum) and the displacement of the bead from the trap center were correlated by a spring constant. The spring constant of the trap was calibrated from the slope of the measured forces vs. the trap bead positions recorded by a CCD camera. The extension of the molecule was controlled by moving the piezoelectric stage, to which a light-lever system was attached to record position changes of the pipette bead. Extension changes of the whole RNA construct were thus obtained from the relative movements of the two beads.

**Hopping experiments**

The folding/unfolding experiments described in this report were performed at ambient temperature (23 ± 2°C) in a buffer containing 10 mM MOPS, pH 7.0, 250 mM NaCl, and 1 mM EDTA. Two types of hopping experiments were done for the P5ab RNA constructs. (A) Constant-force mode (24) (Figure 1C). The force applied to the RNA constructs was maintained at a preset value by moving the piezoelectric stage (pipette bead) through feedback control. Extension of the molecule increased when the RNA unfolded, whereas refolding of the RNA resulted in decrease in extension. The data acquisition rate (bandwidth) for this mode was 200 Hz. (B) Passive mode (Figure 1D). In contrast to the constant-force mode, the piezoelectric stage was left stationary in this mode. Thus, the trap bead can freely move in the trap in response to the end-to-end distance change of the constructs. When the RNA unfolded, the trap bead moved toward the trap center, such that the force decreased; when the RNA refolded, the trap bead was pulled further away from the center to increase the force. Moving the piezoelectric stage to a new position would change the tension on the RNA in folded and unfolded states, such that the equilibrium between these two states would shift. This kind of experiments allowed us to measure kinetics over different forces. The data were collected at 1000 Hz.

**Data analysis**

For the passive mode, folding and unfolding rate coefficients of the P5ab RNA were calculated from the time-dependent force traces. Each trace normally contained at least 50 cycles of unfolding/refolding events, which were usually collected in 10 – 20 s and showed no significant force drift. Distributions of the force were fitted to Gaussian functions for the folding and unfolding processes:

$$y = a_1 e^{-\left(\frac{f-f^U}{c_1}\right)^2} + a_2 e^{-\left(\frac{f-f^F}{c_2}\right)^2}, \qquad (1)$$

where $y$ is the number of counts for each binned force $f$, $a_n$ and $c_n$ ($n$ = 1 or 2) are amplitudes and widths of the peaks, respectively, $f^U$ and $f^F$ are the (average) forces at the unfolded and folded states, respectively, and fluctuations (standard deviations) of the force are $\delta f^U = c_1/\sqrt{2}$ and $\delta f^F = c_2/\sqrt{2}$ (see Figure 4). Under most conditions, $\delta f^U$ and $\delta f^F$ were essentially equal. States (folded or unfolded) of the RNA along the force trace were assigned according to $f^U$ and $f^F$. Rate coefficients were computed as the inverse of the mean lifetime for each state. Alternatively, for a two-state system, the rate coefficients ($k$) can be obtained by fitting the following first-order kinetics equation (14):

$$P = e^{-kt}, \tag{2}$$

where $P$ is the probability that the unfolding or folding reaction has not occurred by the time $t$. Examples of the fitting are shown in Figure 3A. With few exceptions, the rate coefficients calculated from these two methods matched very well.

This simple Gaussian analysis for state assignments works well for two-state systems. For the P5ab hairpin the force distributions (see Fig. 4) showed only two peaks, and no intermediates were detected in our current and previous (20) experiments. Furthermore, modeling of the kinetics by us and others (25), taking into account the breaking and forming of each base pair, predicts no detectable intermediates. However, more complex methods of data analysis could be helpful in investigating the presence of intermediates. For example, McKinney et al. (26) have recently developed an algorithm based on hidden Markov modeling to analyze single-molecule FRET trajectories. This approach makes possible unbiased separation of noise from state-to-state transitions. Because the goal of the present work is to analyze the effect of the experimental setup on the measured folding/unfolding rates (independent of whether intermediates can be detected), we use the simple method of two-state data analysis described by Eq. 1. Similar methods have also been applied for analysis of DNA hairpins (27).

For the constant-force mode, rate coefficients were calculated from the time-dependent extension traces. The data collection usually lasted 3 – 5 min to obtain enough statistical data (usually 60 – 300 unfolding/refolding cycles) for each preset force. As the measured extension traces may drift over the time period (for example, the drift was ~ 10 nm over 30 s in Figure 2A), we applied a different strategy to analyze those data. A transition (unfolding or folding) was considered to occur when the extension was changed by at least 75% of the extension difference (~ 20 nm) of the P5ab RNA between folded and unfolded states under the preset forces (14 – 15 pN). Unfolding and folding rate coefficients were calculated as described above. To obtain the standard deviation in extension, the distribution of extension difference between any two neighboring data points ($\Delta x_i = x_{i+1} - x_i$) was plotted and fitted to a Gaussian function:

$$y = ae^{-\left(\frac{\Delta x - b}{c}\right)^2}, \tag{3}$$

where $b$ is the average of $\Delta x_i$. The fluctuation (standard deviation) of the extension ($x_i$) is $\delta x = (c/\sqrt{2})/\sqrt{2} = c/2$. (Compared to the definition in Eq. 1, the extra factor of $\sqrt{2}$ corrects the uncertainty difference between $\Delta x_i$ and $x_i$ in a Gaussian distribution.) Note that $\Delta x_i$ contains not only fluctuations but also transition signals. Since the folding/unfolding rates (< 10 s$^{-1}$; see Table 1) are much smaller than the data acquisition rate (200 Hz), contribution of transition signals to $\Delta x_i$ is thus not significant.

**Spatial and force resolution**

The amplitude of the fluctuations in extension ($x$) and force ($f$) measured with a bandwidth $B$ is given by the integral over frequency, $\omega$, of the power spectrum of the mean square displacement for a particle in a harmonic potential with effective stiffness $\varepsilon_b + \varepsilon_x$ (28):

$$\langle \delta x_B^2 \rangle = \int_0^{2\pi B} \frac{k_B T}{\varepsilon_b + \varepsilon_x} \left[ \frac{2\omega_c}{\pi(\omega^2 + \omega_c^2)} \right] d\omega \quad \text{and} \quad \langle \delta f_B^2 \rangle = \varepsilon_b^2 \langle \delta x_B^2 \rangle, \qquad (4)$$

where $k_B$ is the Boltzmann constant, $T$ is the temperature of the bath (reaction chamber), $\varepsilon_b$ and $\varepsilon_x$ are the stiffness of the optical trap and the RNA construct, respectively, and $\omega_c$ is the corner frequency of the bead. In the case of infinite bandwidth the fluctuations are given by the equipartition result (29):

$$\langle \delta x^2 \rangle = \frac{k_B T}{\varepsilon_b + \varepsilon_x} \quad \text{and} \quad \langle \delta f^2 \rangle = \frac{\varepsilon_b^2 k_B T}{\varepsilon_b + \varepsilon_x}, \qquad (5)$$

Fluctuation (square roots of Eq. 4) is a measurement of noise, and thus magnitude changes of signals in extension or force smaller than the fluctuation will not be detected by the instrument. In other words, Eq. 4 gives the resolution limits $\Delta x_{RL}$ and $\Delta f_{RL}$ for the extension and force, respectively, at a given bandwidth $B$:

$$\Delta x_{RL} = \sqrt{\langle \delta x_B^2 \rangle} \quad \text{and} \quad \Delta f_{RL} = \sqrt{\langle \delta f_B^2 \rangle}. \qquad (6)$$

To quantify the measurability of a system, we define a signal-to-noise ratio (*SNR*) as the ratio between the amplitude of the signal (changes in extension or force) and the corresponding resolution limit:

$$SNR_x = \frac{\Delta x}{\Delta x_{RL}} \quad \text{and} \quad SNR_f = \frac{\Delta f}{\Delta f_{RL}}. \qquad (7)$$

Theoretically, a dynamic process with *SNR* larger than 1 may be detected and measured.

When the bandwidth is much smaller than the corner frequency of the bead, i.e., $B \ll \omega_c$ (as in the case of constant-force mode), Eq. 4 can be approximated to:

$$\langle \delta x_B^2 \rangle = \frac{4(B/\omega_c) k_B T}{\varepsilon_b + \varepsilon_x} = \Delta x_{RL}^2 \quad \text{and} \quad \langle \delta f_B^2 \rangle = \varepsilon_b^2 \langle \delta x_B^2 \rangle = \Delta f_{RL}^2. \qquad (8)$$

In contrast, the bandwidth effect disappears when $B \gg \omega_c$, and the fluctuations will be given by Eq. 5, which differs from Eq. 8 by a factor of $4(B/\omega_c)$. Therefore, by using a smaller bandwidth (e.g. averaging over a longer period of time) the measured fluctuations can be reduced and the *SNR* increased, resulting in better spatial and force resolution.

**Temporal resolution**

If the data collected from the experiments were instantaneous, the temporal resolution would be limited by the relaxation time of the bead, $\tau_b$ (= $1/(2\pi w_c)$). In practice, the data collected are always averaged over a bandwidth $B$. To measure an event with a given lifetime $\tau$ the bandwidth must satisfy $1/B < \tau$. Thus, in general, the limit of time resolution, $\Delta t_{RL}$, is given by either $1/B$ or $\tau_b$:

$$\Delta t_{RL} = \max\{1/B, \tau_b\}. \qquad (9)$$

For the constant-force mode the time resolution can also be limited by the time lag of the feedback control of the instrument, $T_{lag}$:

$$\Delta t_{RL} = \max\{1/B, \tau_b, T_{lag}\}. \qquad (10)$$

Similarly, we can define a temporal *SNR* as the ratio of the characteristic time ($\tau$) of the dynamic processes, i.e., the lifetime of the folded and unfolded conformers of the RNA, over the resolution limit ($\Delta t_{RL}$):

$$SNR_t = \frac{\tau}{\Delta t_{RL}}. \tag{11}$$

**RESULTS**

Force unfolding of RNA using optical tweezers is induced by pulling the two ends of the RNA (through handles and beads) and by monitoring the changes in force and extension of the whole construct. In the case of P5ab RNA, the unfolding event is characterized by a sudden increase in extension and decrease in force in the force-extension curve (20), as demonstrated in Figure 1B. Inversely, refolding of the RNA is detected by a sudden extension drop and force increase, which usually follows the reverse trace of the unfolding pathway, showing that the force folding/unfolding process is reversible. The unfolding/refolding forces were in the range of 13.5 – 15.5 pN in the buffer system used (pH 7.0, 250 mM NaCl, 1 mM EDTA). When the pulling/relaxation rate was 2.5 pN/s or less, the RNA jumped back and forth between two extension values at forces near 14.5 pN (Figure 1B, inset), indicating a fast structural transition between the folded and unfolded states. This phenomenon is called *hopping* (20).

In this study, the unfolding and folding rates of P5ab were measured using the hopping method, which was carried out in two modes: the constant-force mode and passive mode (see Materials and Methods for details). Since the RNA undergoes cycles of folded and unfolded states under either mode, lifetimes in each state can be measured many times from one single experiment, making hopping a convenient method to study kinetics.

**Handles and pulling experiments**

One of the major factors we changed to study the effects on kinetics of P5ab RNA was the lengths of the double-stranded RNA/DNA handles. Physical properties of the hybrid handles were investigated by pulling experiments, and the force-extension curves were fit to the worm-like chain model (WLC) (30, 31):

$$F = \frac{k_B T}{P} \left[ \frac{1}{4(1-x/L)^2} + \frac{x}{L} - \frac{1}{4} \right], \tag{12}$$

where $F$ is the force, $P$ is the persistent length, $k_B$ is the Boltzmann constant, $T$ is the temperature, $x$ is the extension (end-to-end distance), and $L$ is the contour length. The fitting was applied to the low-force (< 14 pN) region, where the hairpin was still closed and thus could be excluded for the analysis. The end-to-end distance $x$ of the RNA/DNA constructs was calculated from the positions of the pipette bead and trap bead from video images. Note that the light-lever system only records the *extension change*, not the end-to-end distance, of the tether (see Materials and Methods). From the fitting, the persistent length $P$ was 21.7 ± 3.6 nm and the contour length $L$ 0.25 ± 0.01 nm/bp (N = 21) for the 10.2 Kbp construct, and $P$ 12.0 ± 4.0 nm and $L$ 0.26 ± 0.02 nm/bp (N = 17) for the 5.9 Kbp construct. Thus, the apparent persistent length of RNA/DNA hybrids seems to be length-dependent; the shorter the tether, the smaller the persistent length. It is not clear why the measured persistent length changes with tether length, but some possibilities are: (i) the RNA/DNA hybrids are attached to the surface of the polystyrene beads, or are partially melted due to differences in hydration, ionic strength, or pH; (ii) the contour length and the bending and twisting rigidity of the tethers depend on extension (32); (iii) the assumption in the WLC equation that L (the total contour length) >> P is not satisfied for shorter tethers (30). The curves from 1.1 and 3.2 Kbp constructs could not be fit well due to shorter extension changes of these tethers and limitation of the video images. Empirically we find that a persistent length of 10 nm (or smaller) is a good approximation for shorter hybrids. In addition, the contour lengths (0.25 – 0.26 nm/bp) were consistent with the structures of A-form duplexes and RNA/DNA hybrids (0.26 – 0.29 nm/bp) (33-35), supporting the validity of the fitting approach we used here.

For pulling experiments, we found that the size of the rips changed with handle lengths; the extension changes decreased from 16.7 ± 1.4 nm (1.1 Kbp) to 12.7 ± 1.7 nm (10.2 Kbp) and the force changes decreased from 1.0 ± 0.1 pN (1.1 Kbp) to 0.7 ± 0.1 pN (10.2 Kbp). The rip size is also affected by the trap stiffness (28). Thus, the reversible work measured under the rip was significantly reduced from 142.3 ± 12.5 KJ/mol (1.1 Kbp) to 108.7 ± 15.0 KJ/mol (10.2 Kbp). The difference in mechanical work was mainly caused by contraction of the handles when the force dropped in the rip, and longer (softer) handles had larger effects. The energy contribution from the handles can be calculated at the unfolding force (14.5 pN) using the worm-like chain model to obtain the work of unfolding the RNA. By making this correction, the work for the four constructs (1.1, 3.2, 5.9, and 10.2 Kbp handles) falls in the range of 162 – 173 KJ/mol, independent of handle lengths. For a reversible process, the work done to/by the system is equal to the Gibbs free energy change, and thus this subtraction of the handles' effects verifies that our experimental measurements provide the free energy of force unfolding RNA.

Another method to calculate the free energy change is from kinetics measurements (see next section for details). At the critical force the rates for unfolding and folding are equal; the reaction is reversible and the work is equal to the Gibbs free energy. The critical forces ($F_{CFM}^c$) and extension changes ($\Delta x$) were essentially the same for different handles (see Table 1). Thus, the free energy change in this process (179 KJ/mol) was obtained by multiplying the average force (14.5 pN) with the average extension change (20.5 nm). This is consistent with the results of pulling experiments (162 – 173 KJ/mol). Furthermore, the Gibbs free energy change for unfolding the RNA at zero force can be obtained by subtracting the stretching energy of the single-stranded RNA at the unfolding force (44 ± 10 KJ/mol; (20)), and these values, 118 – 129 KJ/mol for pulling and 135 KJ/mol for constant-force experiments, are comparable to the predicted value 138 kJ/mol by *Mfold* (at 37°C, 1 M NaCl; (36, 37)). Therefore, the Gibbs free energy for unfolding the RNA can be correctly measured, independent of handle lengths and kinetic methods.

**Constant-force mode**

The constant-force mode was done by holding the P5ab RNA constructs at a preset tension near the transition force, usually between 14 and 15 pN. The unfolding and folding processes were followed by the extension traces over time. As shown in Figure 2, the extension of the RNA hopped between two distinct sets of values, with the larger one corresponding to the unfolded state and smaller one to the folded state. The difference between the two sets of extension, ~ 20 nm, reflected release of the 49 nucleotides involved in the hairpin structure (see Figure 1A) (20). During hopping, the force fluctuated around the preset value. The amplitude of force fluctuation varied with trap stiffness; it was about 0.4 pN and 0.2 pN for the 0.1 pN/nm and 0.035 pN/nm optical traps, respectively. The extensions of both folded and unfolded conformers may slowly drift at a rate of up to 1 nm/s, largely due to mechanical instability of the chamber and the detector. As we varied the handle lengths and trap stiffness, the extension difference of the two states ($\Delta x$) remained constant (Table 1). Close-up views of the hopping traces are shown in Figures 2B and 2D. An interesting observation was that the force fluctuation was not always stochastic; instead, an extension jump was accompanied with a force burst, which then relaxed to the preset value. Our explanations for this phenomenon are as follows. When the RNA unfolds or folds, the tension between the two ends of the RNA immediately decreases or increases, respectively, to cause the force bursts. Because the force feedback control of the machine, limited by the piezoelectric stage, does not respond as fast as the tension change, a time lag occurs before the force is gradually restored to the preset value. The operation time of the feedback for the current setup is about 0.1 second (see the transitions indicated by asterisks in Figure 2B and 2D). Correspondingly, the force may vary up to 2 pN during this recovery time, such that the probability that the reverse reaction would occur in this period can be significantly different from that at the intended force. For example, a transition in Figure 2B (indicated by an arrow) shows that the RNA hairpin folded at a force 1 – 2 pN lower than the preset value. Because lower forces encouraged RNA molecules to fold, this transition could be induced by the temporarily lowered force and thus should be considered as a folding event at that force. Similar arguments are applicable to unfolding processes. The 0.1 s time lag of the feedback control was consistent for all the P5ab RNA constructs with different handle lengths and optical trap stiffness (data not shown).

Kinetics of P5ab were measured with four different handle lengths (1.1, 3.2, 5.9, and 10.2 Kbp) and two values of trap stiffness (0.1 and 0.035 pN/nm). Unfolding and folding rate coefficients ($k$) were obtained as described in Materials and Methods. The rate coefficient is assumed to depend exponentially on the applied force ($F$) (14, 38, 39):

$$k(F) = k_m k(0) e^{(FX^{\ddagger}/k_B T)}, \tag{13}$$

where $k_m$ reflects the contribution of instrumental factors (including the handles) to the rate (20), $k(0)$ is the rate constant at zero force, $X^{\ddagger}$ is the distance to the transition state, $k_B$ is the Boltzmann constant, and $T$ is the temperature. Thus, a linear relationship is expected from the plot of $\ln(k)$ versus $F$ (for a force range where $X^{\ddagger}$ is constant), as shown in Figure 3B. As the force is increased, the unfolding rate increases and the folding rate decreases. The two curves meet at a point where the rates are equal; this unique rate is called the critical rate $k_{CFM}^c$ and the corresponding force is called the critical force $F_{CFM}^c$, where the subscript CFM stands for Constant-Force Mode. At the critical force (also called $F_{1/2}$ or $F_m$), the RNA has the same tendency to fold and unfold. As mentioned above, transitions with lifetimes less than the time lag (~ 0.1 s) of the feedback control may not occur at the desired force, and it is difficult to accurately assign those transitions to individual forces. Therefore, to simplify the analysis, we treated those short-lifetime transitions (< 0.1 s) as a group and computed two rate coefficients for each measurement: $k_{CFM}^c$(cutoff = 0 s) including all measured transitions and $k_{CFM}^c$(cutoff = 0.1 s) excluding those transitions with lifetimes less than 0.1 s. As discussed above, the short-lifetime transitions are promoted by the temporarily changed forces during the recovering period, and their inclusion will lead to the overestimation of the unfolding or refolding rates. Thus, $k_{CFM}^c$(cutoff = 0 s) will define the upper limit of the critical rates measured under this condition.

On the other hand, $k_{CFM}^c$(cutoff = 0.1 s) excludes all the short-lifetime events, which may include transitions happening at the desired force, and thus it will define the lower limit of the critical rates. In this context, we consider that the two sets of $k_{CFM}^c$ span the error of measurements associated with the feedback control. Results of the constant-force kinetics measurements under different conditions are summarized in Table 1. The critical force $F_{CFM}^c$ and extension changes $\Delta x$ upon transition did not significantly change with optical trap stiffness and handle lengths, showing that the Gibbs free energy change of the folding/unfolding process of the RNA remained the same, independent of the experimental setup (see above). From Table 1, we can rank several factors affecting the measured critical rate coefficients of P5ab under constant-force mode. (i) Optical trap stiffness: the rates increased by 2.4 – 3.4 fold when the trap stiffness was lowered to one third (from 0.1 to 0.035 pN/nm). (ii) Effective bandwidth: by cutting off all the 0.1 s transitions, the bandwidth was effectively reduced from 200 to 10 Hz. The rates at 200 Hz (cutoff= 0 s) were 1.6 – 2.3 fold higher than those at 10 Hz (cutoff = 0.1 s). (iii) Length of the handles: the rates varied slightly (< 40%) when the handle lengths were changed by ~ 10 fold (1.1 to 10.2 Kbp). Therefore, under the current setup and conditions, the optical trap stiffness and bandwidth affected the P5ab kinetics more than the handle length did.

**Passive mode**

In addition to the constant-force mode, kinetics of P5ab was measured using a different type of hopping experiments, called passive mode. The passive mode is operated by leaving the pipette bead stationary (without feedback) and allowing the trap bead to "passively" move in the trap (compare Figures 1C with 1D). A similar experimental design (using two optical traps) has been applied recently to study kinetics of a series of DNA hairpins (27). As the hairpin unfolds, the tension on the whole molecule decreases due to the single-stranded RNA released from the hairpin, whereas folding of the RNA causes the force to increase. As a result, the RNA unfolds at a high force and refolds at a low force; both the force and extension are changed during the structural transition. On the force trace, the folded state was accordingly assigned to the regions with higher forces and the unfolded states to the regions with lower forces, as shown in Figure 4, left panels. These two force regimes were well characterized by force distribution plots, to which Gaussian functions (Eq. 1) can be fitted (Figure 4, right panels). The folding/unfolding forces and force standard deviation were defined from the fitting (see Materials and Methods for details). The force distribution showed only two corresponding peaks for the unfolded and folded states, and no apparent intermediates were detected, consistent with a two-state kinetic system.

Differences between the two force regimes ($\Delta f = f^F - f^U$; see Eq. 1) changed with handle length and trap stiffness; $\Delta f$ became smaller for longer handles and softer traps (see Discussion for more details on $\Delta f$). Under the current conditions, $\Delta f$ varied from 0.53 to 1.46 pN (Table 2). The two force regimes may overlap significantly for small $\Delta f$ and large standard deviations of the force $\delta f$ (related to the width of the peaks; Eq. 1), and thus the boundary between the folded and unfolded states will be uncertain. Measurability of the transition can be quantitatively defined by the signal-to-noise ratio $SNR$, $\Delta f / \delta f$ (Eq. 7; Table 2). The smallest $SNR$ was about 2.8 for the 10.2 Kbp handles and 0.1 pN/nm trap, indicating that even in this case it is possible to detect the structural transitions (see the next section for more details on $SNR$). In practice, in some cases, assignments of transitions on the force trace were sometimes ambiguous due to partial overlapping of the two distribution peaks (Figure 4B), reflecting the fact that we were approaching the resolution limits in this extreme case.

The unfolding and folding rates of the RNA in the passive mode were measured from the high and low force regimes, respectively, as described above. By varying the position of the pipette bead, we recorded a series of passive hopping data at different pairs of forces. As in the constant-force mode, a linear correlation between the force and the logarithm of rate coefficients was observed; the unfolding rates increased and folding rates decreased with force (Figure 3C). Likewise, a critical force $F_{PM}^c$ was defined as the force when both rates were equal, and this unique rate was called $k_{PM}^c$ (Table 2). As in the constant-force mode, $F_{PM}^c$ remained unchanged with different handle lengths and laser trap stiffness, and the values were consistent in both modes (see Tables 1 and 2). The effects of handles and trap stiffness on the critical rates seemed to be comparable. For a given handle length, the rates changed up to 60% with trap stiffness. For a given trap stiffness, the rates changed by a factor of 2 or less with handle length. Overall, the critical rates measured by the passive mode with 1000 Hz bandwidth fell in the same order and in the range of 3 – 7.5 $s^{-1}$, despite the dramatic changes in the experimental setup (10- and 3-fold variations on the handles and trap stiffness).

**Limitation in the kinetics measurements**

Changes in experimental setup may not only affect the kinetics but also their measurability. A better understanding of practical limitations of the measurements (or resolution limits) of the current experiments is necessary to interpret correctly the kinetic data. As shown in Eq. 6, the resolution limits were defined by the standard deviation of the measurements, which were computed with the fluctuation-dissipation theorem (Eq. 4) for the passive mode (at 1000 and 200 Hz) and constant-force mode (at 200 Hz), as shown in Figure 5. The corresponding standard deviations from experimental measurements are also plotted. Note that the experimental standard deviations at 200 Hz for the passive mode were obtained by averaging and recalculating the measured 1000 Hz raw data. In the constant-force mode, the calculations from theory showed that the standard deviation in extension was almost constant (~ 0.8 nm), independent of handle length and trap stiffness, whereas the standard deviation obtained from experiments tended to increase slightly (2.2 to 3.0 nm) with handle length, but remained unchanged with different traps stiffness for a given handle length (Figure 5A). The measured deviations were significantly higher than the theoretical ones; the ratio was about 3.2. For the passive mode at 1000 Hz, the theory predicted that the standard deviation in force was more sensitive to the trap stiffness and short handles (smaller than ~ 1.5 Kbp); the deviation decreased (~0.17 to 0.06 pN) when the trap stiffness was lowered from 0.1 to 0.035 pN/nm, but it did not change with handle length (longer than ~ 1.5 Kbp) (Figure 5B). The standard deviation from measurements also showed dependence on traps but not handles, and the magnitudes were about 2.3-fold higher than the theoretical values. When the bandwidth in the passive mode was reduced to 200 Hz, the ratio between the standard deviations of the experiments and theory was increased to ~ 3.2 (Figure 5C), consistent with the results in the 200 Hz constant-force mode. The fact that a unique constant rescaling factor at a given bandwidth is required to fit the fluctuation-dissipation theorem suggests that there may be common sources of additional, uncorrelated noise associated with the experimental system. In fact, for the 0.1 pN/nm trap, force measurements in a bead immobilized on the tip of the micropipette show low frequency (around 3 Hz) noise with an amplitude of ~ 0.3 pN due to mechanical vibrations (data not shown). This instrumental noise adds to the thermal noise fluctuations of a free bead in the trap held at 15 pN giving a total amplitude noise of $(0.17^2 + 0.30^2)^{0.5} = 0.34$ pN, in good agreement with the standard deviation in the force reported in the experiments (in the range 0.34 – 0.36 pN, see Table 2).

To further investigate how the fluctuations in force or extension affected the measurability of kinetics, we calculated the theoretical (Eq. 7) and experimental (Tables 1 and 2) *SNR*. As shown in Figure 6A, the theory predicted a nearly constant *SNR* value of about 25 for the constant-force mode, independent of handle length (> 1 Kbp) and trap stiffness, whereas the *SNR* values from experiments were consistently smaller, within the range of 6 – 10. Given the fact that the standard deviations (Figure 5A) and extension changes (Table 1) were basically independent of handle length and trap stiffness in the constant-force mode, it was not surprising to see a similar tendency for *SNR*. In the passive mode, the theoretical calculations showed that the *SNR* change with handle length was more dramatic in the short handle region (< 2 Kbp) and relatively moderate with longer handles (Figure 6B). The *SNR* calculated from experimental measurements also showed a modest decrease with handle length and the values were all near or below 5; it dropped by ~ 1/3 when the handle length was increased from 1.1 to 10.2 Kbp (Table 2). As shown previously (Figure 4), this change was significant in this case; the two force regimes from the passive mode (0.1 pN/nm trap) were separated well for 1.1 Kbp handles (*SNR* = ~ 4.4) but partially overlapped for 10.2 Kbp handles (*SNR* = ~2.8). Therefore, the measurability of kinetics will be affected more significantly when the *SNR* is approaching the theoretical threshold (*SNR* = 1).

The temporal resolution was about $10^{-3}$ s for the passive mode at 1000 Hz bandwidth (Eq. 9) and 0.1 s for the constant-force mode (largely reduced due to the feedback control; Eq. 10). In our experimental setup the time resolution might be also reduced by temporal correlations in electronic noise. The measured critical rate coefficients in either mode were not greater than 10 s$^{-1}$ (Tables 1 and 2), i.e., the average lifetime for the folded and unfolded states was longer than 0.1 s. As a result, the temporal signal-to-noise ratio $SNR_t$ (Eq. 11) was greater than 100 for the passive mode, but close to 1 for the constant-force mode. The $SNR_t$ in the constant-force mode indicated that the time resolution limited the ability to faithfully follow the structural transitions of the RNA. Nevertheless, having considered the features of the feedback mechanism and how the RNA responds accordingly, we could estimate the effect of the temporal resolution on the measured rates (by using different values of lifetime cutoffs, see above). The current instrumental setup only limited kinetic measurements in the constant-force mode; the other types of measurements were well within the temporal resolution limit of the instrument.

In general, better time resolution can be achieved by increasing the bandwidth *B* (especially for the passive mode), which, however, can impair the spatial and force resolution (compare figure 5, panels B and C). The balance among spatial, force and temporal resolution should be considered when choosing a proper bandwidth for the system of interest. For example, one may use a wide time averaging window to detect a slow transition with a small spatial signal, whereas for fast hoppers, such as the P5ab RNA, the passive mode with a high bandwidth is a better choice.

**DISCUSSION**

Mechanical force exerted through optical tweezers is a powerful approach to study kinetics of RNA folding/unfolding, particularly for simple RNA hairpins (20, 24). The force can be applied to RNA in at least three different ways: force-ramp (pulling experiments, as shown in Figure 1B), constant-force hopping (as shown in Figures 1C and 2), and force-jump (24). In this work, we introduced another hopping method, passive mode force hopping (see Figures 1D and 4). The measured critical rate coefficients for the P5ab RNA from the constant-force mode were not consistent with those from the passive mode under comparable conditions (see Tables 1 and 2). Factors that cause this discrepancy in rate coefficients can be different in each case, but some of them may play an important role in general, such as the intrinsic property of the RNA and the physical setup of the tweezers. The properties of the RNA that influence the measurable kinetic behavior include: whether the reaction is two-state, or shows intermediates, the range of rates in the reaction, etc. Here we have focused on the instrumental and experimental effects, including application of force, lengths of the handles, and stiffness of the optical trap. In addition, the response time of the feedback control for the constant-force mode can affect the values of the measured kinetic parameters of the RNA folding/unfolding reaction when the hopping rate of the RNA between its folded and unfolded states is faster than the speed at which the feedback system operates. However, as shown above, the measured critical rates for P5ab varied by only 7-fold in the range of 1.2 – 8.7 $s^{-1}$ (see Tables 1 and 2). In general, the measured kinetic parameters of the RNA folding/unfolding reaction were affected only moderately by instrumental setup under the conditions tested here.

**The distortion effect under the constant-force mode**

According to the analysis in the companion paper (1), the relaxation times of the handles ($10^{-8}$ – $10^{-6}$ s) and the beads ($10^{-5}$ – $10^{-3}$ s) in the optical trap are much shorter than the response time of the feedback system (0.1 s). Thus the micropipette does not move as soon as the tension between the two beads changes due to a structural transition of the RNA. Within the response time of the feedback (up to 0.1 s), the change in the RNA extension ($\Delta x_r$) is mainly distributed to both the flanking handles and trapped bead, resulting in the handle contracting/relaxing ($\Delta x_h$) and the bead moving toward/away from the trap center ($\Delta x_b$). In an optical tweezers experiment, only $\Delta x_b$ is measured by the instrument and it reflects the changes in both force and extension. When the optical trap is much softer than the handles ($\varepsilon_b<<\varepsilon_h$), $\Delta x_b$ approaches $\Delta x_r$, i.e., the measured extension changes will reflect the actual RNA hairpin transition distance. At the other extreme when the optical trap is much stiffer than the handles ($\varepsilon_b>>\varepsilon_h$), $\Delta x_h$ approaches $\Delta x_r$ and $\Delta x_b$ approaches 0 (28), i.e., the end-to-end distance of the construct does not change as the RNA folds/unfolds, and thus the RNA transition processes can not be detected. Under most conditions the situation is in between these two extremes: the extension change can be measured, but it is smaller than $\Delta x_r$. We call this the *distortion effect*; we see hopping with a height smaller than $\Delta x_r$ (~ 20 nm for the P5ab RNA) on the extension trace, as demonstrated in Figure 2, panels B and D (indicated by #). As we could not always distinguish those transitions from noise, we empirically set a threshold that only transitions with extension changes greater than 75% of the $\Delta x_r$ value were considered as real transitions (see Materials and Methods).

To examine how important the distortion effect was in our current experimental setup, we calculated the stiffness of the handles using the worm-like chain model, with persistent lengths of 22 and 12 nm for the 10.2 and 5.9 Kbp constructs, respectively, and assuming 10 nm for the 1.1 and 3.2 Kbp constructs (see above). The stiffness of the handles $\varepsilon_h$ falls in the range from 1.17 to 0.19 pN/nm at 14.5 pN (the average transition force) for 1.1 to 10.2 Kbp handles, all greater than that of the optical trap ($\varepsilon_b$ = 0.035 – 0.1 pN/nm). For the 1.1 Kbp handles ($\varepsilon_h$ = 1.19 pN/nm), $\varepsilon_h>>\varepsilon_b$ was basically satisfied (especially for the 0.035 pN/nm trap), and thus the distortion effect was expected to be relatively small. Figures 2A shows an example for the 1.1

Kbp handles with the 0.1 pN/nm trap; except for some short-peak transitions on the right side, most transitions were full length. When the softer 0.035 pN/nm trap was used for the 1.1 Kbp handles, the observed hops corresponded almost exclusively to full-length transitions (not shown), further supporting the conclusion that the distortion effect is not significant in a system with short (stiff) handles and soft optical traps. In contrast, when the 10.2 Kbp handles ($\varepsilon_h = 0.13$ pN/nm) were used in the stiffer trap (0.1 pN/nm), $\varepsilon_h \cong \varepsilon_b$, and the distortion effect became substantial as shown in Figures 2C and 2D.

Comparison of Figures 2B and 2D shows that the distortion effect distorts the square-wave-like extension traces and sometimes makes the transition assignments ambiguous. As shown above, using short handles and soft traps can minimize this effect. In this regard, the softest trap is obtained by placing the bead in the anharmonic region of the trapping potential in which the stiffness of the trap is essentially zero (19). Within this anharmonic region (~ 50 nm) the force remains constant, equivalent to an instantaneous force feedback system. Therefore, the distortion effect should vanish in the zero trap stiffness setup. In addition, based on current constant-force measurements, the kinetic rates increased by 2.4 – 3.4 fold when the trap stiffness was lowered to one third (see Results and Table 1). Thus, the rate is expected to increase if the RNA is placed in the zero-stiffness trap.

**Other options for constant-force measurements**

As mentioned above, the force in our tweezers system is maintained constant through a feedback loop, which can limit measurements of fast transitions. In this respect, magnetic tweezers (13, 40, 41) could be an instrumental alternative for the constant-force mode. A magnetic field can produce a constant, uniform force over a spatial range of centimeters (42). The typical operating force is in the range of 0.01 – 10 pN, but can be increased to 20 pN or higher, which includes the unfolding force of P5ab (~ 14.5 pN). However, the position of the magnetic bead is usually tracked by video images that have low temporal resolution for tracking distance changes and low spatial resolution (~ 10 nm; (40)), although the resolution could be improved by tracking the magnetic bead using a low-power laser beam (43). Also, short tethers are a problem for magnetic beads because the tethers attach at a specific magnetic latitude on the bead (44), which causes the tether to wind partially around the bead when the external field orients the bead. Thus, magnetic tweezers are most useful for constructs with long handles and large transition distances. For those RNA or DNA structures with smaller transitions, an option is to use the two-trap optical tweezers having an essentially constant force region spanning ~ 50 nm (19) (see above).

**Correlation of effective stiffness and rates**

In our current system, the P5ab RNA folding/unfolding rates were affected more significantly by the optical trap stiffness than by the handle length, especially for the constant-force mode. The measured rate coefficients changed moderately when the handle length increased by a factor of 10 (from 1.1 to 10.2 Kbp) for a given trap stiffness, whereas the change was similar or larger when the trap stiffness varied by only 3-fold (from 0.1 to 0.035 pN/nm) for any given handle length (see Tables 1 and 2). In this context, we may consider that it is the effective stiffness (optical trap + handles) and not the individual ones, what is more important in affecting the measured kinetics. The stiffness of the optical trap $\varepsilon_b$ (0.035 and 0.1 pN/nm) was always smaller than that of the handles $\varepsilon_h$ (1.17 to 0.19 pN/nm for 1.1 to 10.2 Kbp handles; see above), and thus the former would largely dominate the effective (combined) stiffness $\varepsilon_{eff} = \varepsilon_b \varepsilon_h / (\varepsilon_b + \varepsilon_h) \cong \varepsilon_b$, when $\varepsilon_b << \varepsilon_h$ (especially for short handles and/or soft traps). The effective stiffness qualitatively explains why the measured kinetics was influenced by the optical trap more significantly than by the handle length for the current experimental setup. However, from a quantitative point of view, the measured critical rate coefficients did not always correlate well with the effective stiffness. Figure 7A shows the critical rate coefficients of the P5ab RNA as a function of effective stiffness in the constant-force and passive modes. The rates measured at the 0.035 pN/nm trap are clustered in a small stiffness region of the figure (left side, below 0.04 pN/nm) and are mostly larger than those measured at the 0.1 pN/nm trap (towards the right side) for each mode, but it is unlikely that each data set can be correlated by a simple, common function. Therefore, the effective stiffness of the system is not the sole factor to affect the folding/unfolding kinetics of the RNA molecule.

**Unique features of the passive mode**

The passive mode applied in this work has some unique features; the main difference from other modes is that the force in the passive mode is not controlled. This feature can introduce experimental complications when a long-time measurement is required, because the instability of the physical setup (such as the micropipette) can cause significant drift. In the present study, the drift was not systematic but random and less than 0.1 pN/s. At present, this mode is only suitable for fast RNA hoppers, such as P5ab (typically displaying ~ 60 cycles of folding/unfolding processes in ~ 10 s).

In passive mode, the force makes transitions between two force regimes. These two force regimes are well defined by Gaussian functions (see Figure 4). The force distribution also follows a Gaussian function when a folded or unfolded state predominates by adjusting the force far away from the transition force (e.g., at 10 or 20 pN; data not shown). These results suggest that, in these experiments, the force applied to the molecule is maintained to within a narrow distribution of values before a structural transition happens, even though it is not being actively controlled. This feature allows us to assume that the unfolding reaction occurs at one "constant" force and the folding reaction occurs at the other, making the determination of force-dependent critical rates possible (as in the constant-force mode). On the other hand, the force difference ($\Delta f$) measured in this passive mode changed with handle length significantly (1.46 – 0.96 pN) in the 0.1 pN/nm trap but only moderately in the 0.035 pN/nm trap (0.65 – 0.53 pN; see Table 2). As discussed above, the bead in the trap and handles relax in less than $10^{-3}$ s, which is faster than the hopping rates of the RNA in the passive mode. Thus, the bead movement in the trap (proportional to $\Delta f$) upon a transition should be related to the effective stiffness $\varepsilon_{eff}$ of the system. In the companion paper, we predict a linear relationship between $\Delta f$ and $\varepsilon_{eff}$ (Eq. 3.3 in reference (1)). As can be seen in Fig. 7B, such linear relationship is indeed observed experimentally ($R^2 =$ 0.985). Thus, the value of $\Delta f$ in the force hopping passive mode, is much more predictable than the value of the critical rates in that mode.

Finally, in the passive mode, RNA unfolding occurs at one force and refolding at another, in a way reminiscent of force-jump experiments (24). In the force-jump mode, the force is initially held at a value far from the transition force before rapidly stepping to a new value, and then the unfolding or folding event is monitored. The force is maintained through a feedback system as in the constant-force mode. After the transition takes place, the force is rapidly reset to another value to monitor the reverse transition, and this procedure is repeated. The design of force-jump experiments allows measurements of kinetics in a much wider range of forces. Note that how the force is controlled and manipulated makes the passive, constant-force, and force-jump modes different from each other. This difference is likely to affect the values of the critical rates obtained with these various modes, but the values of the critical forces remain constant.

**Conclusions**

The measured kinetics of the P5ab RNA hairpin in our current optical tweezers system fell in the same order (1.2 – 8.7 $s^{-1}$ at the critical force) despite dramatic changes in the experimental setup, including 3-fold difference in optical trap stiffness, 10-fold difference in handle length, 100-fold difference in effective bandwidth, and two modes of force application on the RNA. A recent study on a series of DNA hairpins shows that the kinetic rates can change by several orders of magnitude when varying the stem-loop sizes and base compositions (27). Thus, it is encouraging that instrumental factors only change the rates to a limited extent. We therefore conclude that optical tweezers are a robust system for studying kinetics of RNA and DNA structures; the variation in kinetics originating from the machinery is relatively small compared to the intrinsic properties of the nucleic acid itself. It is important to understand what experimental designs allow the measurement of a rate approaching the intrinsic molecular rate of an RNA molecule. By combining the experimental results obtained here with simulation studies of the accompanying paper (1), it is possible to deduce the intrinsic molecular rates of the P5ab RNA hairpin and choose the instrumental setup most suitable for such measurements.

Table 1. Critical rate coefficients of P5ab RNA in constant-force mode (200 Hz bandwidth)

| Handle length (Kbp) | $F_{CFM}^{c}$ (pN) | $\Delta x$ (nm) | $k_{CFM}^{c}$ (s$^{-1}$), cutoff= 0 s | $k_{CFM}^{c}$ (s$^{-1}$), cutoff= 0.1 s | $\delta x$ (nm) | $SNR_x$ |
|---|---|---|---|---|---|---|
| Optical trap stiffness = 0.1 pN/nm | | | | | | |
| 1.1 | 14.5 ± 0.3 | 20.5 ± 0.7 | 2.60 ± 0.25 | 1.24 ± 0.09 | 2.2 ± 0.2 | 9.5 ± 0.6 |
| 3.2 | 14.3 ± 0.1 | 19.8 ± 0.2 | 2.78 ± 0.23 | 1.40 ± 0.08 | 2.3 ± 0.2 | 8.6 ± 0.7 |
| 5.9 | 14.5 ± 0.0 | 20.4 ± 0.0 | 3.00 ± 0.27 | 1.53 ± 0.11 | 2.8 ± 0.0 | 7.2 ± 0.1 |
| 10.2 | 14.5 ± 0.0 | 20.4 ± 0.4 | 2.48 ± 0.24 | 1.59 ± 0.18 | 2.9 ± 0.5 | 7.2 ± 1.1 |
| Optical trap stiffness = 0.035 pN/nm | | | | | | |
| 1.1 | 14.6 ± 0.1 | 21.1 ± 0.3 | 6.19 ± 0.65 | 3.09 ± 0.25 | 2.4 ± 0.2 | 9.0 ± 0.7 |
| 3.2 | 14.7 ± 0.2 | 20.6 ± 0.7 | 8.68 ± 0.76 | 3.72 ± 0.30 | 2.5 ± 0.2 | 8.2 ± 0.5 |
| 5.9 | 14.7 ± 0.4 | 21.2 ± 0.1 | 8.63 ± 0.13 | 3.75 ± 0.08 | 2.9 ± 0.2 | 7.5 ± 0.5 |
| 10.2 | 14.6 ± 0.1 | 20.0 ± 0.6 | 8.44 ± 0.43 | 3.84 ± 0.16 | 3.0 ± 0.1 | 6.7 ± 0.1 |

Table 2. Critical rate coefficients of P5ab RNA in passive mode (1000 Hz bandwidth)

| Handle length (Kbp) | $F_{PM}^c$ (pN) | $\Delta f$ (pN) | $k_{PM}^c$ (s$^{-1}$) | $\delta f$ (pN) (unfolding) | $SNR_f$ (unfolding) |
|---|---|---|---|---|---|
| Optical trap stiffness = 0.1 pN/nm | | | | | |
| 1.1 | 14.7 ± 0.3 | 1.46 ± 0.09 | 2.95 ± 0.46 | 0.34 ± 0.05 | 4.4 ± 0.9 |
| 3.2 | 14.0 ± 0.4 | 1.31 ± 0.06 | 4.99 ± 0.19 | 0.36 ± 0.03 | 3.6 ± 0.5 |
| 5.9 | 14.3 ± 0.5 | 1.15 ± 0.02 | 5.27 ± 0.52 | 0.36 ± 0.04 | 3.2 ± 0.4 |
| 10.2 | 14.3 ± 0.4 | 0.96 ± 0.10 | 6.05 ± 0.89 | 0.34 ± 0.01 | 2.8 ± 0.3 |
| Optical trap stiffness = 0.035 pN/nm | | | | | |
| 1.1 | 14.7 ± 0.1 | 0.65 ± 0.01 | 4.62 ± 0.29 | 0.13 ± 0.01 | 5.0 ± 0.3 |
| 3.2 | 14.6 ± 0.2 | 0.60 ± 0.03 | 6.38 ± 0.44 | 0.14 ± 0.01 | 4.4 ± 0.6 |
| 5.9 | 14.7 ± 0.3 | 0.59 ± 0.01 | 6.36 ± 0.34 | 0.15 ± 0.01 | 3.9 ± 0.1 |
| 10.2 | 14.7 ± 0.1 | 0.53 ± 0.01 | 7.53 ± 0.58 | 0.16 ± 0.01 | 3.4 ± 0.2 |

**FIGURE LEGENDS**

**Figure 1**: (A) The single-molecule construct. The P5ab RNA sequence is shown in the hairpin structure, which is flanked by hybrid RNA/DNA handles. The handles A and B are tagged with biotin and digoxigenin molecules at the ends, which are bound to polystyrene beads coated with streptavidin and anti-digoxigenin antibody, respectively, as shown in the bottom panels. (B) Force-extension curves of the RNA construct with 1.1 Kbp handles. The RNA is pulled (black) and relaxed (gray) at a loading rate of ~ 2.3 pN/s. Note that these two traces basically overlap. Inset, detail of the force-extension curves showing that RNA hops between the folded and unfolded states at forces around 14.5 pN. (C) Hopping experiments in the constant-force mode. The RNA is placed between two beads, one (the top bead, 3 μm in diameter) held in the laser trap and the other (the bottom bead, 2 μm in diameter) fixed to the tip of a micropipette by suction. The micropipette is moved up or down to compensate extension changes on the RNA undergoing structural transitions, such that the tension on the RNA (i.e., the position of the bead in the trap) is maintained. The pipette movement is controlled by a feedback loop using a proportional, integration and differentiation algorithm. (D) Hopping experiments in the passive mode. The micropipette does not move in this mode. The trap bead moves toward the trap center when the RNA unfolds, such that the force decreases; refolding of the RNA causes the trap bead moving away from the trap center to increase the force. Drawings in panels (A), (C) and (D) are schematic and not to scale.

**Figure 2**: Time-dependent extension and force traces of the P5ab RNA in the constant-force mode (with the 0.1 pN/nm trap) for 1.1 Kbp (A) and 10.2 Kbp (C) handles. (B) and (D), corresponding zoom-in regions from (A) and (C), indicated by rectangle windows. Examples showing the delay (~ 0.1 s) of the feedback control have asterisks. The arrow in (B) shows an example that the transition occurs at a force different from the preset value (14.5 pN). Transitions showing the distortion effect are indicated by pound signs (#). U: unfolded states, F: folded states.

**Figure 3**: (A) Plots of the probability of P5ab of the unfolded (○) or folded state (●) as a function of time. This set of data was measured in the passive mode with 1.1 Kbp handles and 0.035 pN/nm trap. The forces on the unfolded and folded states were 14.3 and 14.9 pN, respectively. The data were divided into 25 bins and fitted to exponential decay functions (solid curves; Eq. 2); the folding and unfolding rate coefficients are respectively 6.5 and 16.8 $s^{-1}$. Plots of ln($k$) as a function of force for the constant-force mode (B) and passive mode (C). Panel (B), 5.9 Kbp handles and 0.1 pN/nm trap; (C), 1.1 Kbp handles and 0.035 pN/nm trap (same as panel A). The unfolding (●) and folding (○) rate coefficients of the P5ab RNA increase and decrease with the force, respectively. Linear regression curves (solid lines) are fitted to the data for each state, and the critical forces and rate coefficients are obtained from the crossing point of the two lines. Each circle represents one measurement, which contains 150 – 300 transitions for panel (B) and 40 – 80 transitions for panel (C).

**Figure 4**: Force traces and distribution of the P5ab RNA in the passive mode. Examples are shown for the P5ab RNA constructs with 1.1 (A) and 10.2 Kbp (B) handles in 0.1 pN/nm trap. The force distribution is fitted to Gaussian functions, from which the unfolding and folding forces are defined by the peaks (right panels). The transition forces are used as thresholds to assign the folded or unfolded states on the force traces (left panels, thick lines). U: unfolded states, F: folded states.

**Figure 5**: Fluctuations as a function of handle length for the 0.1 pN/nm (blue circles or lines) and 0.035 pN/nm (red circles or lines) traps. Circles and solid lines are the data obtained respectively from experimental measurements and theoretical calculation according to Eq. 4. Dashed lines are obtained from the solid lines by multiplying by a constant factor (see below) to match the measured data. (A) Extension fluctuations ($\delta x$) from the constant-force mode with a bandwidth of 200 Hz (Table 1). The correction factor is 3.2. Note that the red and blue lines overlap. (B) Force fluctuations ($\delta f$) in the passive mode with a bandwidth of 1000 Hz (Table 2). The correction factor is 2.3. (C) Force fluctuations ($\delta f$) in the passive mode with an averaged bandwidth of 200 Hz (see text). The correction factor is 3.2, the same as in (A).

**Figure 6**: Signal-to-noise ratios of (A) extension for the constant-force mode ($SNR_x$; Table 1) and (B) force for the passive mode ($SNR_f$; Table 2) as a function of handle length. Circles and lines are the data obtained respectively from experimental measurements and theoretical calculation according to Eq. 7. Filled and open circles are for 0.1 and 0.035 pN/nm traps, respectively; solid and dotted lines are for 0.1 and 0.035 pN/nm traps, respectively. Note that the two lines mostly overlap in panel (A).

**Figure 7**: (A) Correlation of critical rate coefficients and effective stiffness (trap plus handles). The effective stiffness is calculated from the equation: $\varepsilon_{eff} = \varepsilon_b \varepsilon_h / (\varepsilon_b + \varepsilon_h)$, in which $\varepsilon_b$ and $\varepsilon_h$ are the stiffness of the bead in the trap and the handles, respectively. Shown are data from the constant-force mode at 200 Hz with 0.1 s cutoff (open squares), passive mode at 1000 Hz (filled circles), and passive mode at 200 Hz (filled triangles). (B) Linear relationship between the force changes ($\Delta f$) in the passive mode (at 1000 Hz, see Table 2) and effective stiffness. Data are fitted to linear regression ($R^2 = 0.985$).

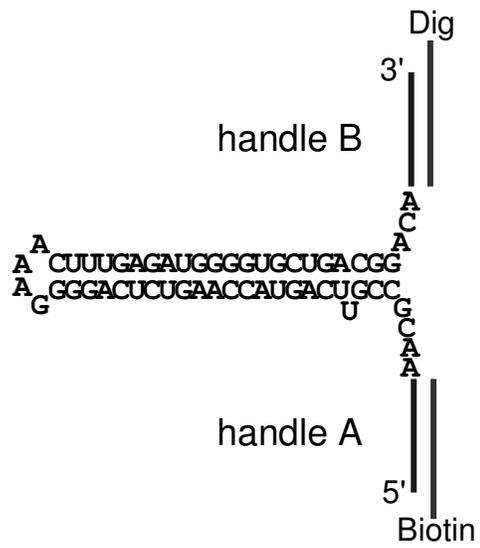
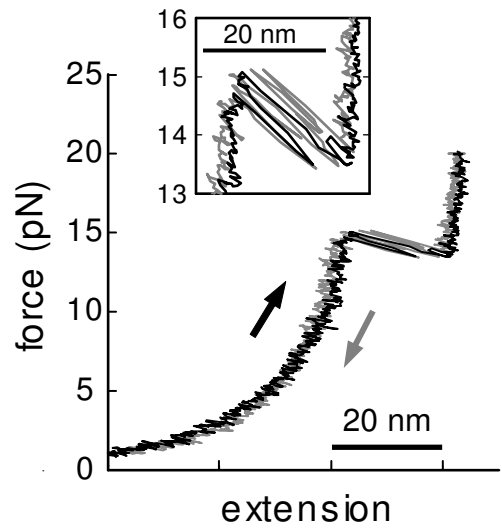
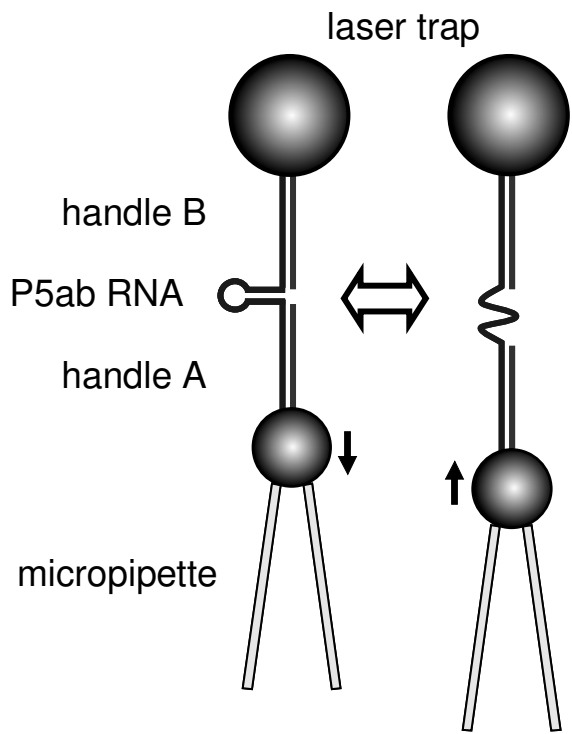
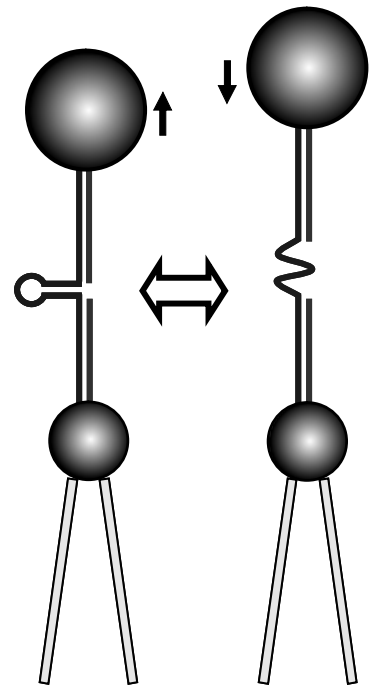

Fig. 1

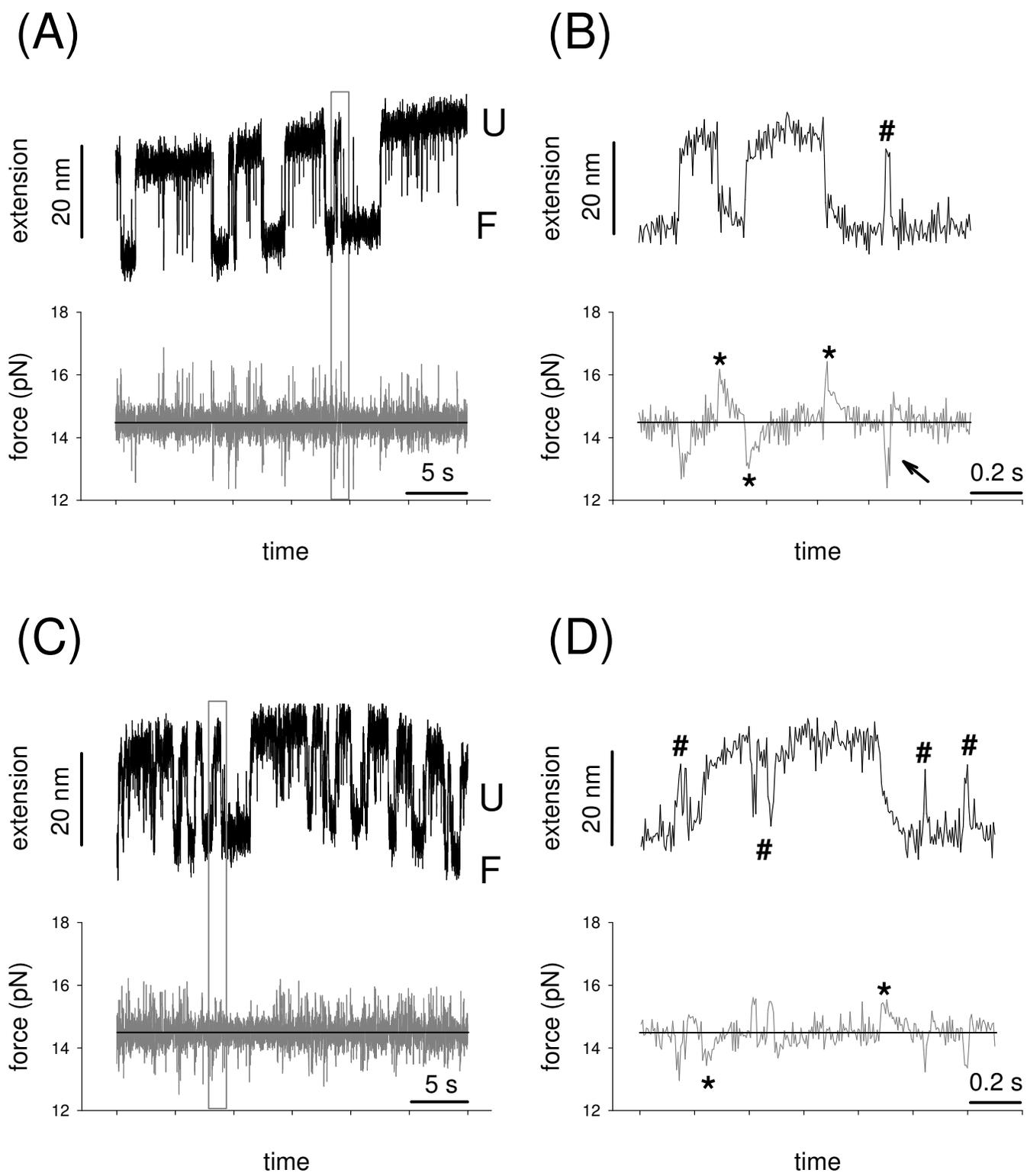

Fig. 2

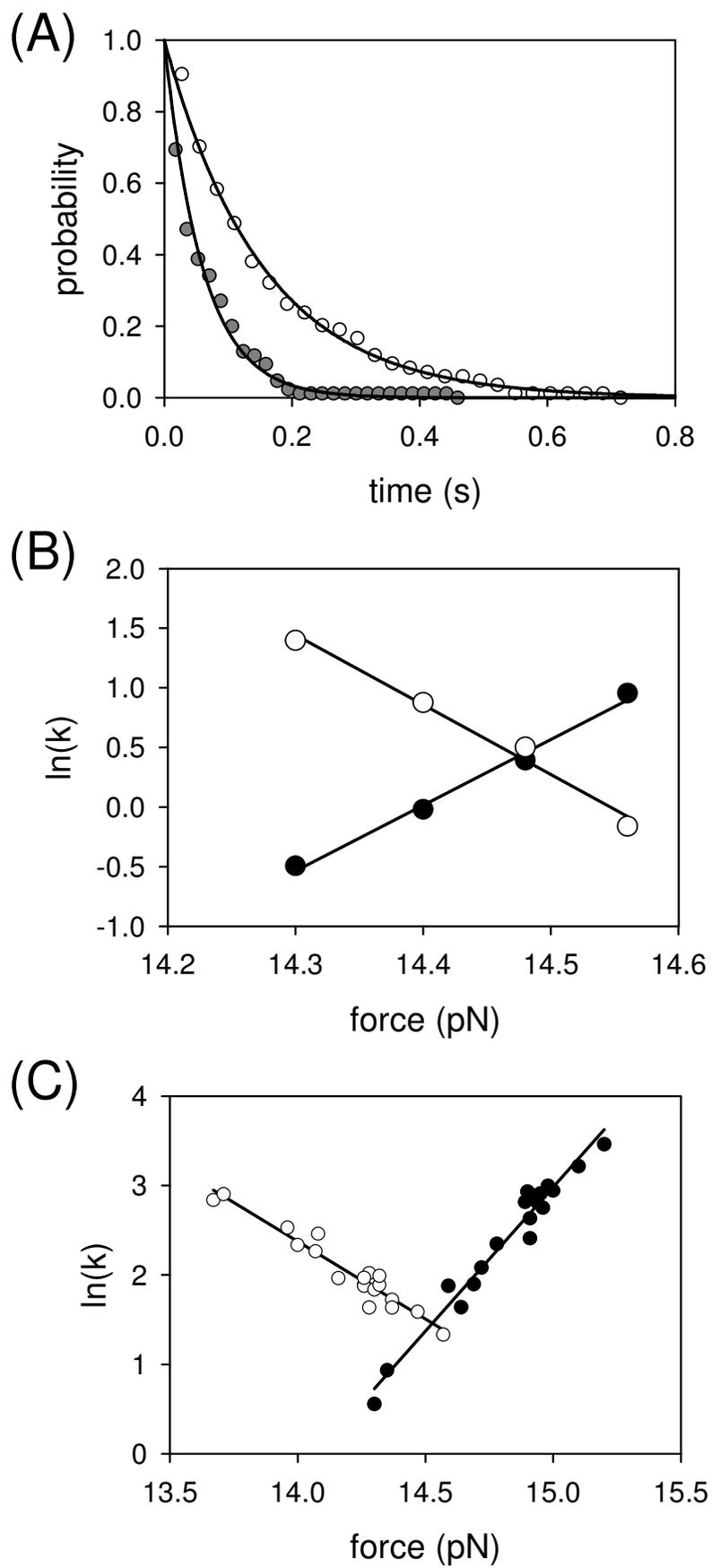

Fig. 3

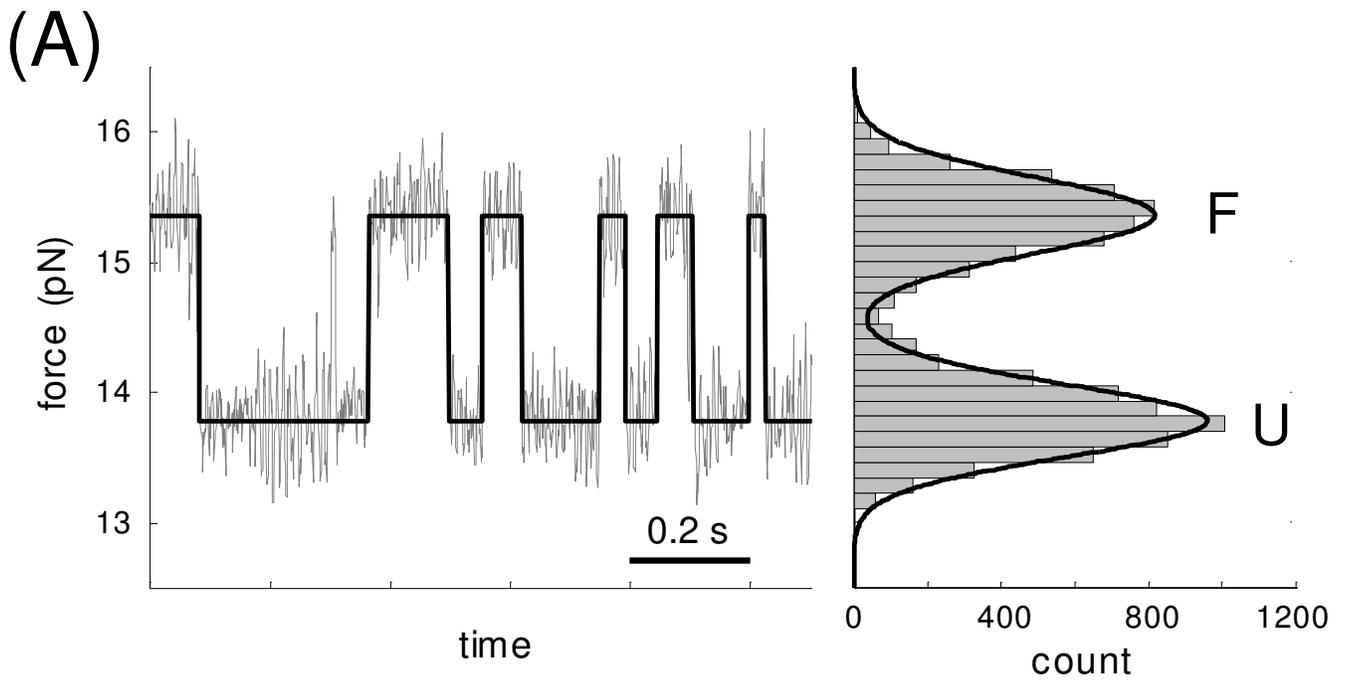

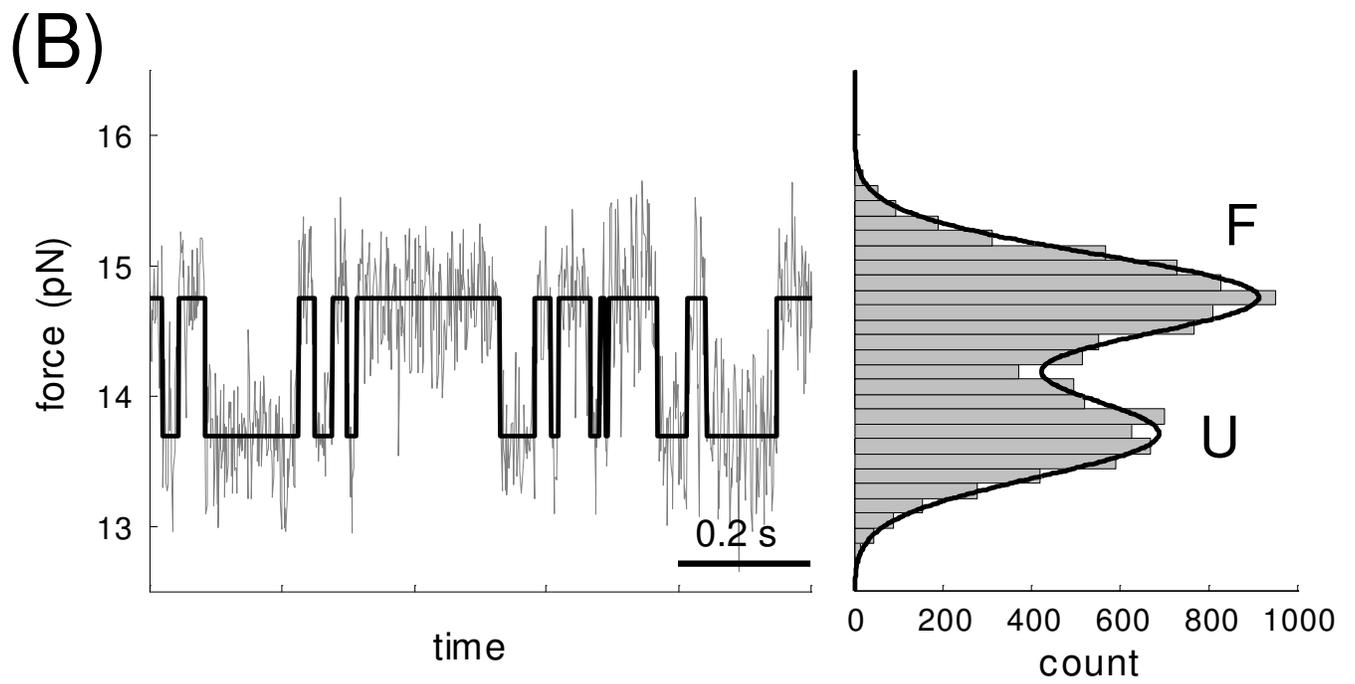

Fig. 4

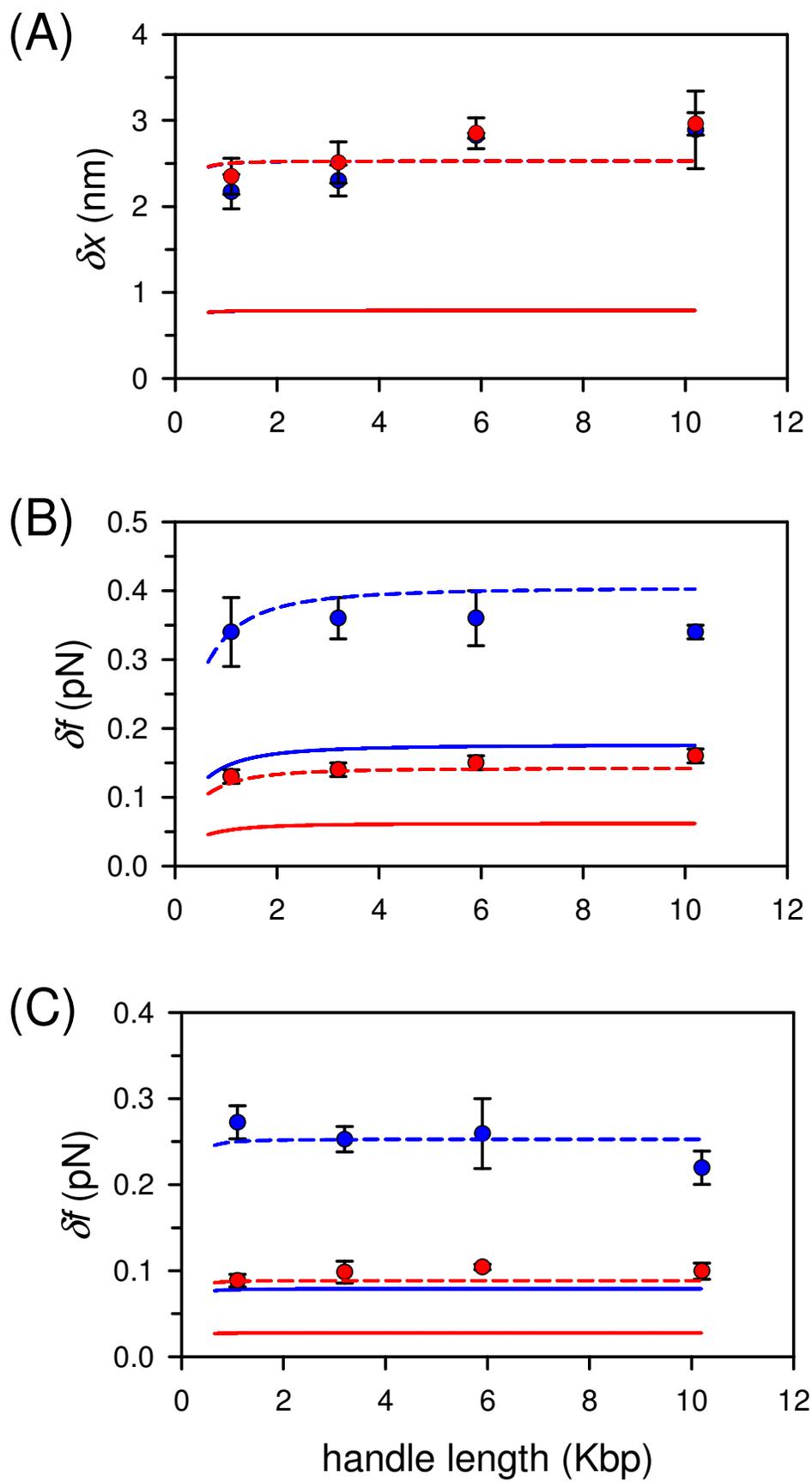

Fig. 5

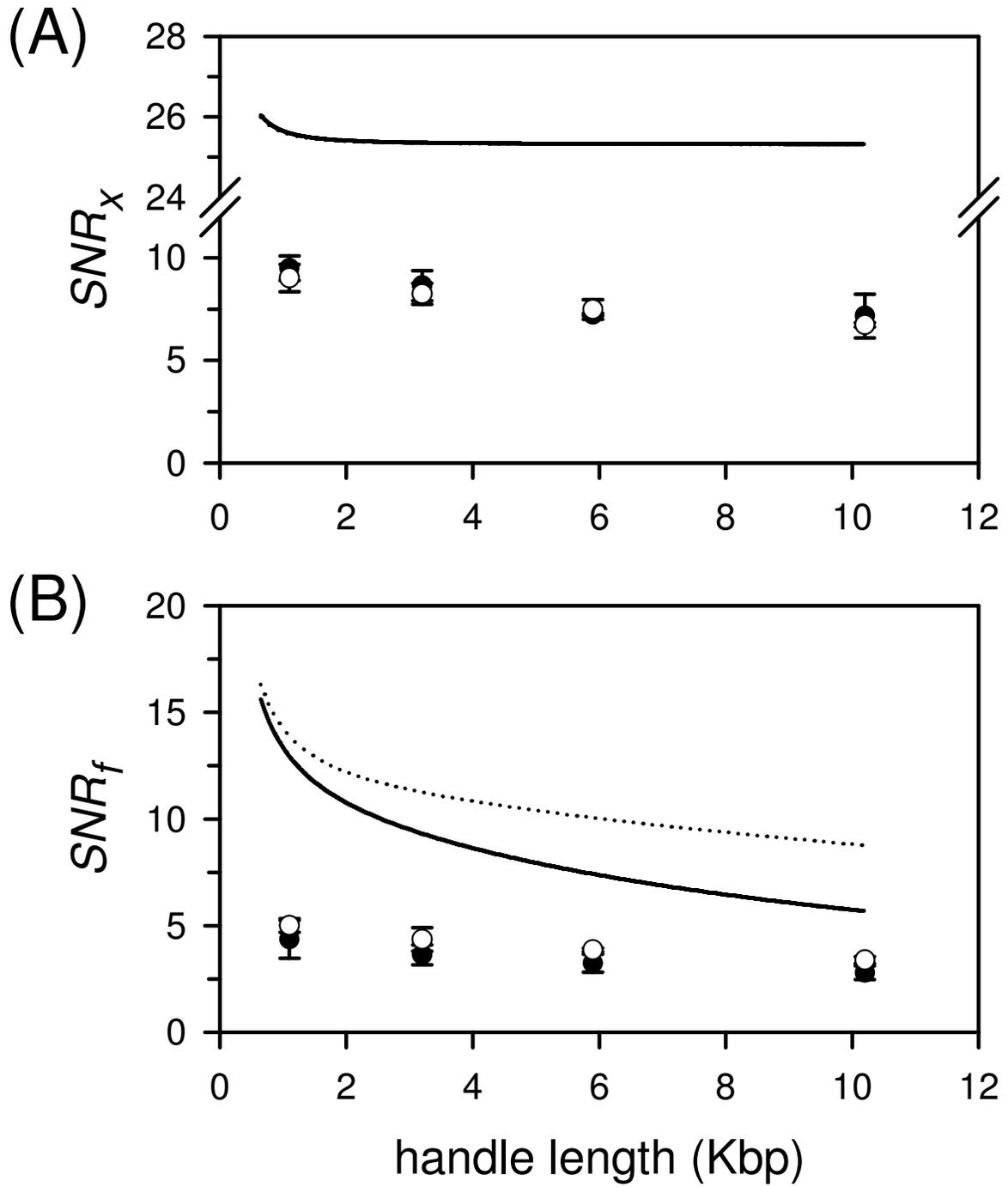

Fig. 6

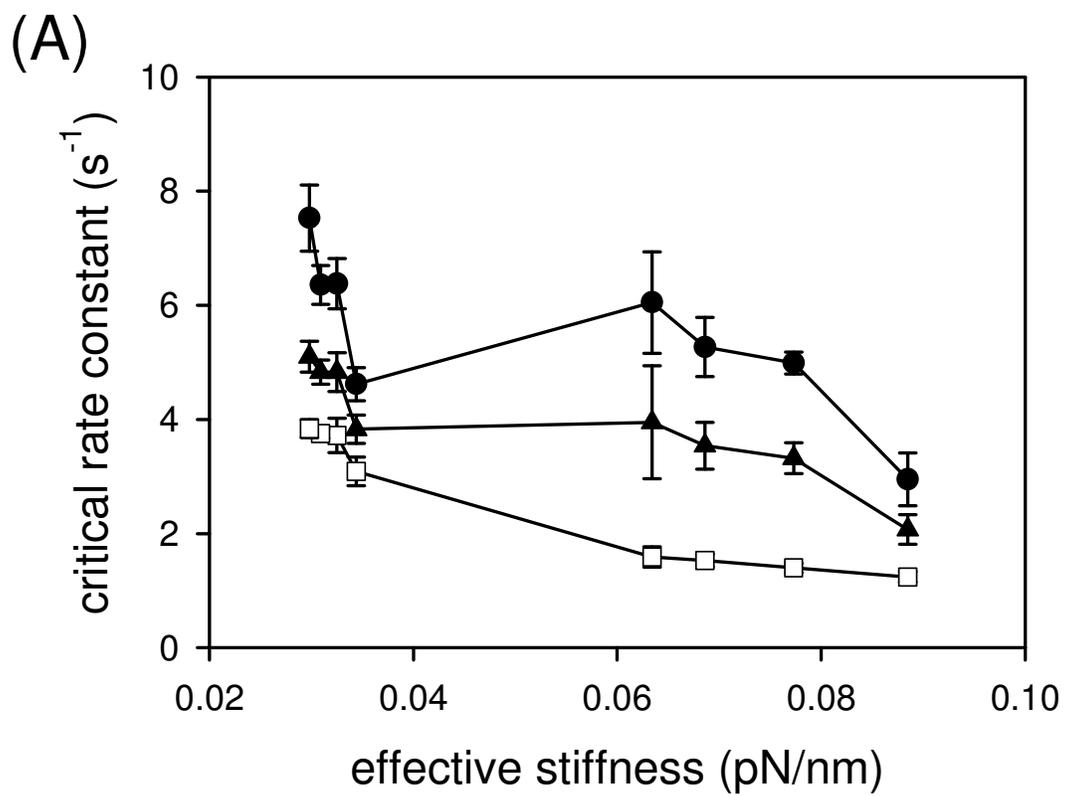

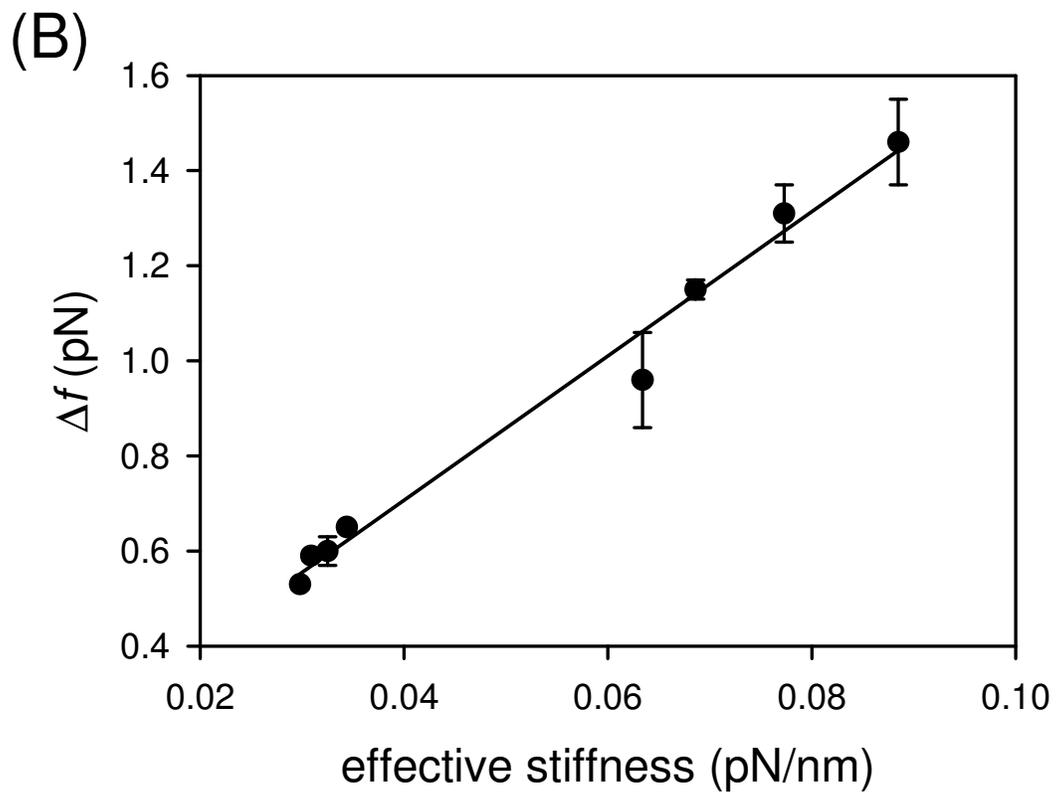

Fig. 7